\documentclass[aps,twocolumn,floats,prd,nofootinbib,10pt,longbibliography,superscriptaddress,amsmath,amssymb]{revtex4-1}
\pdfoutput=1

\usepackage{graphicx}
\usepackage{dcolumn}
\usepackage{bm}
\usepackage[breaklinks=true,colorlinks=true,linkcolor=blue,urlcolor=blue,citecolor=blue]{hyperref}
\usepackage{multirow}
\usepackage[dvipsnames]{xcolor}
\usepackage[pscoord]{eso-pic}
\usepackage[normalem]{ulem}
\usepackage{slashed}
\usepackage{makecell}
\usepackage{comment}

\newcommand{\placetextbox}[3]{
	\setbox0=\hbox{#3}
	\AddToShipoutPictureFG*{
		\put(\LenToUnit{#1\paperwidth},\LenToUnit{#2\paperheight}){\vtop{{\null}\makebox[0pt][c]{#3}}}
	}
}

\def\be{\begin{equation}}
\def\ee{\end{equation}}
\def\ar{~~~\Rightarrow~~~}

\def\aap{A\&A}
\def\apj{ApJ}

\def\prd{Phys. Rev. D}         
\def\prl{Phys. Rev. Lett.}     

\def\baas{Bulletin of the AAS}
\def\sovast{Soviet Astronomy} 
\def\actaa{Acta Astronomica}

\begin{document}

\placetextbox{0.88}{0.985}{\small ULB-TH/23-01}
\placetextbox{0.875}{0.97}{\small CCTP-2023-4}
\placetextbox{0.875}{0.955}{\small ITCP-2023/4}

\title{Quasi-extremal primordial black holes are a viable dark matter candidate}

\author{Jose A. de Freitas Pacheco}
\affiliation{Universit\'e de la C\^ote d'Azur - Observatoire de la C\^ote d'Azur, Bd de l'Observatoire, 06304 Nice Cedex, France}
\author{Elias Kiritsis}
\affiliation{Universit\'{e} Paris Cit\'{e}, CNRS, Astroparticule et Cosmologie, F-75013 Paris, France}
\affiliation{Crete Center for Theoretical Physics, Institute for Theoretical and Computational Physics, Department of Physics, P.O. Box 2208, University of Crete, 70013, Heraklion, Greece}
\author{Matteo Lucca}
\affiliation{Service de Physique Th\'{e}orique, Universit\'{e} Libre de Bruxelles, C.P. 225, B-1050 Brussels, Belgium}
\author{Joseph Silk}
\affiliation{Institut d’Astrophysique de Paris (UMR7095: CNRS \& UPMC- Sorbonne Universities), F-75014, Paris, France}
\affiliation{Department of Physics and Astronomy, The Johns Hopkins University Homewood Campus, Baltimore, MD 21218, USA}
\affiliation{BIPAC, Department of Physics, University of Oxford, Keble Road, Oxford OX1 3RH, UK}

\begin{abstract}
Black hole evaporation is generally considered inevitable for low-mass black holes, yet there is no confirmation of this remarkable hypothesis. Here, we propose a phenomenological model that appeals to the possible survival of light quasi-extremal primordial black holes as a significant dark matter component and show that the related cosmological and astrophysical constraints disappear for reasonable degrees of quasi-extremality. The results obtained are general, conservative and should be taken as a proof of principle for future, model-specific analyses.
\end{abstract}

\maketitle

\section{Introduction}
Since first postulated in the late `60s \cite{Zeldovich1967Hypothesis, Hawking1971Gravitationally, carr:pbhsformation, chapline:pbhformation}, primordial black holes (PBHs) are regarded as a viable dark matter (DM) candidate, albeit over a highly constrained mass range (see e.g., \cite{Sasaki2018Primordial, Carr2020Constraints, Carr2020Primordial, Villanueva-Domingo2021Brief}). The most appealing example is given by a range roughly between $10^{17}$ and $10^{22}$~g, bounded from below by constraints coming from the impact of the energy injection following Hawking evaporation of the PBHs \cite{Hawking1974Black} on observables such as cosmic microwave background (CMB) anisotropies~\cite{Poulin2017Cosmological, Stocker2018Exotic, Lucca2019Synergy, Acharya2019CMB}, cosmic rays~\cite{Boudaud:2018hqb}, and 21 cm lines \cite{Clark:2018ghm}.

These bounds on PBH evaporation are, however, derived under the assumption that the PBHs are non-rotating and neutral, which maximizes their evaporation rate. In fact, it is well known that if the BHs were charged and/or spinning, their Hawking temperature would decrease, and consequently so would their mass loss rate and luminosity. In the limit where the temperature approaches  zero one has what are referred to as quasi-extremal PBHs (qPBHs). Since increasing degrees of quasi-extremality can be reached, this implies that the aforementioned mass range could be extended to lower masses by decreasing Hawking evaporation.

Nevertheless, the formation and survivability of qPBHs is clearly a contentious issue. The acquisition of extreme spin-to-mass ratios is astrophysically forbidden for BH growth by accretion in thin \cite{1974ApJ...191..507T}  or thick disks~\cite{1980AcA....30...35A}. Even if the PBHs had a spin at formation, it would be lost more efficiently than its mass \cite{1976PhRvD..14.3260P, Unal:2023yxt}, making quasi-extremality impossible to obtain over extended periods of time. A similar discussion also applies to the case of charged BHs~\cite{carter1974charge}.

One can, however, invoke other processes to justify the existence of qPBHs at lower masses. For instance, accretion and Schwinger pair production can reduce the BH charge $Q$, but Hawking evaporation can counter these effects and augment the charge-to-mass ratio \cite{Jacobson:1997ge}. Such relics have often been considered as DM candidates \cite{MacGibbon:1987my, Chen:2004ft, Lehmann:2019zgt, Bai:2019zcd}. Indeed, small quantized BHs have a fundamental stable state defined by a mass equal to the Planck value and a spin $J = \hbar$ \cite{2020PhRvD.101h3022P}. We emphasize furthermore that PBHs can form deep in the radiation era where stabilized qPBHs are possible DM candidates if their fractional abundance at formation is as low as $\sim 10^{-25}$, and the application of extreme value statistics leads to the  expectation that some of the DM may plausibly include a qPBH component~\cite{Chongchitnan:2021ehn}. The motivation for the rarity of extremely light qPBHs comes from the equilibrium charge distribution~\cite{Page1977Particle} once the emission of charged particles of random sign is included, $P(Q)\sim \exp[ {-4\pi\alpha(Q/e)^2}]$, where $\alpha$ is the electromagnetic coupling and the PBHs rms charge satisfies $Q/e=1/\sqrt {8\pi\alpha} \approx 6$. Also PBHs living in an higher-dimensional space are DM candidates~\cite{2022PhRvD.105j3508F, Anchordoqui:2022txe}. Provided that the scale of the extra dimension is of the order of the gravitational radius, we show below that such BHs may be quasi-extremal. Furthermore, as shown in \cite{Dvali:2020wft}, even Schwarzschild BHs may depart from Hawking evaporation after the Page time due to the so-called memory burden effect. If this is the case then the BH evolution is non-Markovian and the evaporation rate is suppressed.

More generally, we argue that since Hawking radiation has never been tested experimentally or even unambiguously demonstrated to occur from a rigorous theoretical perspective, one can treat the approach to extremality as a phenomenological constraint on the Hawking temperature and the associated Hawking evaporation rate. The argument that there are no robust mechanisms for generating quasi-extremality is fallacious in that it rests on our ignorance, and most notably we provide several counterexamples above. We show that useful insights into different types of PBHs are provided by testing the degree of quasi-extremality with cosmological and astrophysical data.

It is therefore not unreasonable to expect qPBHs to exist in a realistic cosmological scenario. Yet, the impact that such quasi-extremality would have on the commonly imposed cosmological and astrophysical constraints  on PBH evaporation has not been systematically considered in the literature so far. Here we address this task and propose a very general phenomenological analysis to be taken as a proof of principle for future, more specific studies. As a result, in our simplified and yet conservative scenario, we find that for values of the quasi-extremality parameter $\varepsilon$ (defined in terms of e.g., the PBH charge $Q$ and mass $M$ as $\varepsilon^2=1-Q^2/M^2$) lower than $\lesssim10^{-3}$ all cosmological and astrophysical constraints allow for PBHs to make up for the totality of the DM, and that they become stable over cosmological times for masses as low as $\sim10^{11}$~g. This implies that for sufficiently low values of $\varepsilon$, qPBHs are a viable DM candidate.

Moreover, another reason for being interested in the possibility that qPBHs could be observable is recent progress in string theory and quantum gravity, suggesting that extremal and nearly-extremal BHs could be more exotic objects than previously thought. For (near-)extremal BHs, although one can take the classical limit in gravity, $M_P\to\infty$, there is a special class of quantum gravity effects that do not decouple \cite{Maldacena:2016hyu, Heydeman:2020hhw, Lin:2022rzw}. Therefore, (near-)extremal BHs are probably the best magnifying lenses that one can use to observe quantum gravitational effects. Hence possibilities for observing them are truly exciting.

This work is organized as follows. In Sec. \ref{sec: theory} we discuss both how such levels of quasi-extremality can be reached and maintained over cosmic times, and how they affect the standard picture of Hawking evaporation. In Sec.~\ref{sec: observables} we overview the many cosmological and astrophysical observables affected by the presence of this quasi-extremality and suggest simple ways to recast existing bounds in the quasi-extremal limit. In Sec.~\ref{sec: results} we present the resulting constraints on qPBHs as a function of both the quasi-extremality parameter and the fractional abundance of the PBHs. In Sec. \ref{sec: conclusions} we conclude with a summary and closing remarks.

\section{Quasi-extremal primordial black holes}\label{sec: theory}
In this section we provide a brief overview of scenarios that can generate quasi-extremal BHs. We do so by highlighting the general features that they share and how the analysis presented here can be considered as a proof of principle for other, more specific examples. We furthermore also discuss how this general scenario would affect the standard picture of BH evaporation.

\subsection{Representative examples}\label{sec: examples}
A first well-known example of qPBHs is given by highly charged BHs, in particular Reissner-Nordstrom (RN) BHs. In this case, to parameterize the degree of quasi-extremality we can define the parameter $\varepsilon$ such that\footnote{Here and henceforth we assume $G=c=\hbar=k_B=1$.}
\begin{equation}\label{Q_to_eps}
    1-\frac{Q^2}{M^2}=\varepsilon^2 \ll 1\,,
\end{equation}
where $M$ is the mass and $Q$ is the electric charge of the BH. RN BHs have two horizons: the outer, corresponding to the event horizon, and the inner, the Cauchy horizon, defined by the zeros of the lapse function, that is
\begin{equation}
    r_{\pm} = M\left[1 \pm \sqrt{1 - \frac{Q^2}{M^2}}\right]=M(1\pm\varepsilon)\,.
\end{equation}
The size of the outer horizon should never be smaller than the BH-associated Compton wavelength. This condition gives the Planck mass as the smallest PBH mass.

The notion of the dual horizon also applies to Kerr BHs, which represent a second possible qPBH solution should they spin to a high degree. Here the similar horizon structure is now characterized by
\begin{equation}\label{horizon}
    r_{\pm} = M\left[1 \pm \sqrt{1 - \frac{a^2}{M^2}}\right]=M(1\pm\varepsilon)\,,
\end{equation}
where $a$ is the dimensionless spin parameter (related to the angular momentum $J$ via $a=J/M$), and, analogously to Eq. (\ref{Q_to_eps}), we can then also define
\begin{equation}\label{a_to_eps}
    1-\frac{a^2}{M^2}=\varepsilon^2\,,
\end{equation}
which shows a similar mass dependence of the relation between $\varepsilon$ and the model-specific parameters $Q$ and $a$. It is therefore clear that much of the model dependence of the aforementioned scenarios can be captured by the single parameter $\varepsilon$, which represents the degree of extremality of the BH. In the following discussion we will then only make use of $\varepsilon$ so as to be as general as possible in our conclusions.

In a realistic cosmological scenario, however, it is well known that any charge or spin would be lost very quickly by any BH population of primordial origin. Quasi-extremal primordial RN or Kerr BHs are therefore not expected to exist today. Based on these models, it is nevertheless possible to envision a scenario where the BH is charged, but instead of standard electromagnetism (EM) the BH is charged under a generic EM-like dark charge whose carriers are always much heavier than the temperature of the BH \cite{Bai:2019zcd}. In this way, one obtains the same mathematical setup as for a RN BH, but with the difference that the charge $Q$ does not get evaporated away from the BH and remains therefore constant.

From Eq. \eqref{Q_to_eps}, however, one can infer that a constant $Q$ does not necessarily imply a constant $\varepsilon$. In fact, since initially the charge will always be smaller than the mass of the BH, $\varepsilon$ will always be larger than zero and the BH will radiate its mass at the rate discussed in the following section. Nevertheless, the smaller the mass (i.e., the more the BH evaporates) the more $\varepsilon$ will approach the zero value and the slower the mass loss rate becomes. This means that a constant charge with $Q<M$ leads to a BH that naturally approaches extremality over cosmic times.

On the other hand, Eq. \eqref{Q_to_eps} shows that if both $Q$ and $M$ are radiated away at the same rate, $\varepsilon$ stays constant. This means that in a setup where the initial charge-to-mass ratio is very close to unity and both quantities get radiated at the same rate, the BH can maintain extremality indefinitely. However, contrary to the constant-charge example mentioned above, this scenario would require a larger degree of fine-tuning, as one would need to fix the characteristics of the charged dark particles to be exactly such that the BH loses charge and mass at the same rate.

Another possibility to obtain (and maintain) quasi-extremality that does not depend on charge or spin but that presents very similar features in terms of $\varepsilon$ is given by higher dimensional BHs \cite{2022PhRvD.105j3508F, Anchordoqui:2022txe}. One could, in fact, consider the 5D gravitational theory compactified on a circle of radius $R$ with $R\lesssim 1\, \mu$m, although the notion of quasi-extremality is generalizable to $d$ dimensions. In that case, BHs with a horizon size smaller than $R$ behave as 5D BHs while those with size larger then $R$ as 4D BHs. One can then show (see App. \ref{app: det_d}) that the evaporation of all Kaluza-Klein (KK) modes of a $d$-dimensional BH leads to an effective  degree of extremality
\begin{equation}
    \varepsilon_{\rm eff}^{-4}={3{(d-3)\over (d-1)}{S_4\over S_d}\left({2\pi R\over r_s}\right)^{2{(d-4)}}}\,.
\end{equation}
A ``large"  BH with $r_s\gg R$ evaporates following the 4D decay equation until its horizon becomes $r_s\simeq  R$. From that point on, it decays as a higher-dimensional BH at a rate that is slower than in 4D. This implies that a higher-dimensional BH would tend to quasi-extremality the more it evaporates, qualitatively just like a constant-charge BH.  

In summary, in the discussion above, we have highlighted different ways to justify the existence of qPBHs which can all be phenomenologically described by the evolution of $\varepsilon$ in the respective scenarios. For simplicity, hereafter we will solely focus on the aforementioned (RN BH) case with a constant $\varepsilon$ value. With respect to the other possibilities, this choice is conservative in the sense that, for the same initial value of $\varepsilon$, in the other scenarios the degree of extremality would only increase and hence they would be covered by the results obtained for the constant $\varepsilon$ case (see \cite{Lehmann:2019zgt,
Bai:2019zcd, 2022PhRvD.105j3508F, Anchordoqui:2022txe} for related but relatively limited discussions). Furthermore, we point out that, although based on a slightly less realistic scenario, ours has to be taken as a useful proof of principle to be applied to more specific examples in the future.

\newpage

\subsection{Evaporation}\label{sec: evaporation}
Given these possible sources of quasi-extremality, it is interesting to consider how one might observe the decay of these quasi-extremal BHs and set constraints on their modified evaporation emission.

The key quantity that is modified by the quasi-extremality of the BHs is their evaporation temperature $T$, which now reads (assuming e.g., a RN BH, but with no loss of generality)
\begin{equation}\label{eq: T}
    T=\frac{1}{4 \pi r_+}\left(1-\frac{Q^2}{r_+^2}\right)=\frac{1}{8 \pi M}\frac{2^2\, \varepsilon}{(1+\varepsilon)^2}\,,
\end{equation}
where $\varepsilon$ encapsulates the deviations from the standard Hawking temperature. Once the temperature is defined, it becomes possible to determine the luminosity $L$ of the BH via the Stefan-Boltzmann black body formula\footnote{Near extremality the Stefan-Boltzmann formula is modified by important grey-body factors that tend to suppress emission (see e.g., \cite{Bai:2023hpd} for a related discussion). We do not consider such factors and therefore our formulae should be considered as upper bounds on Hawking emission near extremality
in the RN case. In the higher-dimensional case such factors are not important.},
\begin{equation}\label{luminosity}
    L =A\sigma T^4\propto r_+^2T^4 \propto \frac{2^6\, \varepsilon^4}{(1+\varepsilon)^6 M^2} \,,
\end{equation}
where $A=4\pi r_+^2$ is the area of the BH and $\sigma$ is the Stefan-Boltzmann constant. This simple dependence of the luminosity is however strictly speaking only valid as long as the BH evaporates at a constant rate. Nevertheless, since different particles can be emitted at different BH temperatures, it is more convenient to interpret the luminosity as the energy emission rate with explicit dependence on the mass loss rate d$M$/d$t$, such that
\begin{equation}\label{luminosity 2}
    L = -\frac{\text{d}M}{\text{d}t}\frac{2^6\, \varepsilon^4}{(1+\varepsilon)^6}\,,
\end{equation}
where the $\varepsilon$ dependence needs to be introduced for consistency with Eq. (\ref{luminosity}).

The mass loss rate of an evaporating BH is commonly defined in terms of the total energy carried away by the emitted particles (due to energy conservation arguments), i.e., following the relation
\begin{align}
    \frac{\text{d}M}{\text{d}t} = - \int \sum_j \frac{\text{d}N_j}{\text{d}t\text{d}E} E \text{d}E\,,
\end{align}
where $\text{d}N_j/\text{d}t\text{d}E$ is the number of emitted particles $j$ of spin $s$ in the energy interval between ($E$, $E+\text{d}E$) and is defined as
\begin{align}\label{eq: dN}
    \frac{\text{d}N_j}{\text{d}t\text{d}E} = \frac{1}{2\pi}\frac{\Gamma_j}{e^{(E-\mu_j)/T}-(-1)^{2s_j}}\,.
\end{align}
Here, $\Gamma_j$ is the dimensionless absorption probability of the given emitted species, which, in full generality, depends on both $M$ and $\varepsilon$. The presence of charge and spin can in fact enhance or reduce the probability of charged particles or particles with spin (mis-)aligned with that of the BH to be emitted from the system. For standard RN BHs charged under EM, \cite{Page1977Particle} found that the impact of the charge on $\Gamma_j$ is of the order of a few percent. In Eq.~\eqref{eq: dN}, $\mu_j$ refers to the chemical potential of a given emitted particle and generally depends on $\varepsilon$. For instance, in the case of standard charged BHs, it would take the form $\mu_j \propto q_j\sqrt{(1-\varepsilon)/(1+\varepsilon)}$, where $q_j$ is the charge of the particle. In our simplified scenario, however, we assume that the particles carrying the charge of the quasi-extremal BH are not emitted from the BH at all (or at least very slowly) and we can therefore neglect these $\varepsilon$-dependent contributions to $\Gamma_j$ and $\mu_j$.

Under this simplifying assumption, the only way $\varepsilon$ affects the mass loss rate is via the exponential dependence of $\text{d}N_j/\text{d}t\text{d}Q$ on the BH temperature $T$. This determines what particles are kinematically available at a given temperature $T$ and it therefore makes sense for it to be dependent on the temperature of the system only, regardless of the characteristics of the BH reaching that temperature. One can then simply extend the validity of the results found in \cite{Macgibbon1990QuarkI, Macgibbon1990QuarkII} according to which
\begin{equation}
    \frac{\text{d}M}{\text{d}t} = -5.34\times10^{25} \frac{f(T)}{M^2} \,\, \text{g/s}\,,
\end{equation}
where $f(T)$ defines the number of emitted species and can be expressed as in Eq. (9) of \cite{Macgibbon1990QuarkII} (see also \cite{Stocker2018Exotic, Lucca2019Synergy} for additional details, updated coefficients and contributions 

\noindent beyond the QCD phase transition).\footnote{Concretely, focusing for instance on the left panel of Fig. 10 of~\cite{Lucca2019Synergy} for a graphical representation, the only aspect of the plot that would be modified by the presence of a non-zero $\varepsilon$ would be the relation between the two horizontal axis, reporting the PBH mass $M$ and the corresponding $T$ values, with the latter being shifted more and more to the left the higher the value of $\varepsilon$.}

With this definition of the mass loss rate it is then possible to compute the lifetime of the BH by integrating Eq.~(\ref{luminosity 2}). Following again \cite{Macgibbon1990QuarkII}, one obtains
\begin{equation}\label{eq: t_ev}
    t_{\rm ev}=6.24\times 10^{-27} \frac{M^3}{f(T)}\frac{(1+\varepsilon)^6}{2^6\,\varepsilon^4} \text{ s}\,.
\end{equation}

\section{Impact on the observables}\label{sec: observables}
Once the mass loss rate due to the PBH evaporation and the related luminosity have been defined, it is possible to analyse how the emission of particles from the PBH affects various cosmological and astrophysical observables such as the CMB anisotropies as well as the cosmic and $\gamma$-ray spectra. Since all of these probes are sensitive to different epochs of the universe, they also constrain different mass ranges, allowing us to cover a wide region of parameter space. In the following sections, we describe all of the constraints and explain how they are affected by the presence of evaporating qPBHs.

\subsection{BBN}\label{sec: BBN}
The first observable we focus on is Big Bang Nucleosynthesis (BBN), which covers the period of light-element formation, such as deuterium and helium, in the early universe  \cite{Cyburt:2015mya}. The predictions of standard BBN are in extremely good agreement with measurements of the corresponding abundances in the first galaxies, where galactic dynamics and star formation have not had the time to affect the primordial abundances yet \cite{10.1093/ptep/ptaa104}. Inferred quantities such as the baryon energy density, the baryon-to-photon ratio and the primordial helium abundance are also consistent with CMB measurements \cite{Aghanim2018PlanckVI}.

Given the success of the standard BBN model, this probe has been often employed to constrain beyond-the-standard-model (BSM) physics, such as  annihilating or decaying DM \cite{Jedamzik:2009uy, Hufnagel:2020nxa} or PBH evaporation \cite{Carr2010New, Acharya2019CMB}, typically delivering the most stringent constraints on these types of models prior to recombination. In fact, BBN can constrain BSM models in a variety of ways, from the impact that they might have on the expansion of the universe (changing e.g., the number of relativistic degrees of freedom) to the photo-disintegration of the light elements after BBN is completed.

Precisely this richness of constraints, however, prevents us from deriving simple and general limits that can be recast for any value of $\varepsilon$. This is due to the fact that, for instance, $\varepsilon$ affects both the overall and the relative amount of injected species (via the modification to $f(T)$) as well as the lifetime of the PBHs. Therefore, different aspects of the standard BBN picture might be modified in non-trivial ways, affecting the magnitude and shape of the constraints. BBN bounds on qPBH evaporation would then have to be derived with dedicated analyses.

For this reason, the accurate inclusion of the BBN constraints in the following discussion goes beyond the proof-of-principle type of study conducted here and will not be considered any further. We note, however, that, albeit the scaling of the constraints is not  directly proportional to $\varepsilon$ as for the probes discussed below, we do expect a significant suppression of the constraints the lower the value is of $\varepsilon$ and that the results of this work will not be affected by the non-inclusion of the BBN constraints.

\subsection{CMB}\label{sec: CMB}
\subsubsection{CMB anisotropies}\label{sec: CMB ani}
CMB anisotropies are very well known to be affected by exotic energy injections during the dark ages (see e.g.,~\cite{Poulin2017Cosmological} for a thorough discussion). In fact, in that period of the thermal history of the universe, the cosmic medium was almost perfectly neutral, allowing the CMB photons to travel straight from the last scattering surface (at $z\simeq 1100$) to us. Any injection of particles with enough energy to ionize the abundant hydrogen atoms would have increased the amount of free electrons, thereby enhancing the probability of further scattering of the CMB photons. This modification of the so-called visibility function would in turn affect the shape of the CMB anisotropy power spectra (both temperature and polarization). Since the observed spectra are in perfect agreement with the $\Lambda$CDM model in the absence of any energy injection \cite{Aghanim2018PlanckVI}, the CMB anisotropies can be used to constrain processes such as the evaporation of PBHs.

In order to estimate the extent to which PBH evaporation affects the CMB anisotropies one needs to determine the energy injection rate, which in this case is given by
\begin{equation}
    \left.\frac{\text{d}E}{\text{d}t\text{d}V}\right|_{\rm inj}=\rho_{\rm cdm} f_{\rm PBH} \frac{L}{M}\,,
\end{equation}
where $f_{\rm PBH}$ is the (primordial) fractional abundance of PBHs with respect to the DM. This injected energy does not, however, necessarily coincide with the effectively deposited energy. In fact, for instance, part of the injected energy might be in form of non-electromagnetically interacting particles and not all of it is spent to ionize the medium (some of this energy would heat  up or excite the plasma). These contributions are commonly taken into account by deposition efficiency $f_{\rm eff}$ and deposition fraction per channel $\chi_c$, respectively \cite{Galli2013Systematic, Slatyer2015IndirectI, Slatyer2015IndirectII, Poulin2017Cosmological, Stocker2018Exotic, Lucca2019Synergy}. Explicitly, this implies
\begin{align}\label{eq: E_dep}
    \left.\frac{\text{d}E}{\text{d}t\text{d}V}\right|_{\text{dep}, c} = \left.\frac{\text{d}E}{\text{d}t\text{d}V}\right|_{\text{inj}} f_{\rm eff}\, \chi_c\,.
\end{align}

A graphical representation of the heating rate due to PBH evaporation is displayed in Fig. 5 of \cite{Lucca2019Synergy}. The consequent impact of PBH evaporation on the free electron fraction can be seen in e.g., Fig. 6 of \cite{Poulin2017Cosmological}\footnote{From the left panel of the figure it becomes clear that heating and ionization rates are rather correlated, so that the heating rate shown in \cite{Lucca2019Synergy} is also indicative of the ionization rate.}, while that in relation to the CMB power spectra can be found in Fig.~6 of \cite{Stocker2018Exotic}. Some important remarks can be drawn from the figures. First of all, the majority of the energy injection takes place around the lifetime of the PBH, similarly to the DM decay scenario. This means that the injection time can be roughly approximated to coincide with the lifetime of the PBHs. Secondly, only PBHs with masses larger than $10^{13}$ g evaporate after recombination and can therefore be constrained with CMB anisotropies.

Based on 
Sec. \ref{sec: evaporation}, the energy deposition rate has a dependence on the PBH parameters of the form
\begin{align}\label{eq: E_prop_PBH}
    \left.\frac{\text{d}E}{\text{d}t\text{d}V}\right|_{\text{dep}} \propto f_{\rm PBH} \frac{f(T)\,2^6\,\varepsilon^4}{M^3\,(1+\varepsilon)^6}\,.
\end{align}
In the $\varepsilon=1$ case, $f(T)$ is almost constant for masses above $10^{13}$ g (it varies at most by a factor 3, see e.g., Fig.~10 of \cite{Lucca2019Synergy}) meaning that the energy deposition rate has a simple dependence of the form $f_{\rm PBH}/M^3$. This implies a proportionality of the constraints on the PBH abundance as $M^3$, which is perfectly recovered in the bounds shown in e.g.,  \cite{Stocker2018Exotic, Lucca2019Synergy, Acharya2019CMB}.

On the other hand, the simple proportionality of Eq.~\eqref{eq: E_prop_PBH} also allows us to take into account the contribution of $\varepsilon$ by simply rescaling the bounds on $f_{\rm PBH}$ by a factor $f(T_{\varepsilon=1})(1+\varepsilon)^6/(f(T)2^6\varepsilon^4)$, where $f(T_{\varepsilon=1})$ is the value of $f(T)$ in the $\varepsilon=1$ case (i.e., for a Schwarzschild BH). The overall order of magnitude of the rescaling is given by the chosen value of $\varepsilon$, with the $f(T)$ ratio introducing a further enhancement of at most the order of a few (and never more than ten). At the same time, since the constraints are time-sensitive -- in this case, for instance, they are most stringent for PBH masses corresponding to $t_{\rm ev}$ around the time of recombination -- also the PBH mass needs to be rescaled. Explicitly, qPBHs survive longer the lower the value of $\varepsilon$. Since the evaporation time $t_{\rm ev}$ is proportional to the PBH mass, this implies that for a qPBH with $\epsilon<1$ and a PBH with $\varepsilon=1$ to evaporate at the same time one needs to assume a lower mass in the qPBH case. The proportionality factor is the one given in Eq. \eqref{eq: t_ev}, i.e., the same as for $f_{\rm PBH}$ above. The combination of the rescalings of $f_{\rm PBH}$ and $M$ enables us to recast existing constraints, such as the ones derived in e.g.,~\cite{Stocker2018Exotic, Lucca2019Synergy, Acharya2019CMB}, for any value of $\varepsilon$.

Here we rely on the bounds derived in \cite{Acharya2019CMB}, which are based on Planck 2015 data \cite{Ade2015PlanckXIII}. Corresponding constraints employing Planck 2018 data \cite{Aghanim2018PlanckVI} have been derived in \cite{Lucca2019Synergy} and seem to be approximately one order of magnitude more constraining than those of \cite{Acharya2019CMB}. Nevertheless, \cite{Lucca2019Synergy} employed a simplified thermal history and made use of a mock likelihood instead of real data. For this reason, and for sake of being conservative, we choose to focus on the results of \cite{Acharya2019CMB}. Furthermore, compared to other results based on Planck 2015 data such as \cite{Stocker2018Exotic, Poulter2019CMB}, the findings of \cite{Acharya2019CMB} overlap well for PBH masses corresponding to lifetimes longer than recombination but improve upon them at lower masses, where the PBHs evaporate before recombination. This is due to a better analysis of the delay between the energy injection and its deposition which extends the constraints down to evaporation redshifts of the order of $z\simeq5\times 10^3$.

\subsubsection{CMB spectral distortions}\label{sec: CMB spec}
In brief, CMB spectral distortions (SDs) are any type of deviation of the CMB energy spectrum from a pure black body \cite{Chluba2011Evolution, Lucca2019Synergy, Chluba2019Spectral, Chluba2019Voyage}. They are typically created by the injection of energy or photons in the thermal bath, although they can also be produced by effects such as the dissipation of acoustic waves and adiabatic cooling. Complementary to the CMB anisotropies, CMB SDs are very sensitive to the thermal history of the universe prior to recombination, up to redshifts of the order of ${z\simeq2\times10^6}$.

In the context of PBH evaporation, as in the case of the CMB anisotropies, their shape is determined by the amount of injected energy defined in Eq. \eqref{eq: E_dep}. A key difference is, however, that for CMB SDs it is the heating rate that needs to be considered and not the ionization rate (which would anyway be zero before recombination). This implies that the same rescaling of existing bounds (see e.g., \cite{Tashiro2008Constraints, Lucca2019Synergy, Acharya:2019xla, Chluba2020Thermalization}) discussed in the previous section can be employed for SDs as well. Here we follow the results of~\cite{Acharya:2019xla} based on FIRAS data \cite{Fixsen1996Cosmic,Bianchini:2022dqh}, which perfectly overlap with the more recent and exact calculations of~\cite{Chluba2020Thermalization} at very high evaporation redshifts (or, equivalently, for very low PBH masses), but extend them until recombination.

\subsection{21 cm}\label{sec: 21}
Another cosmological observable that can be employed to constrain the evaporation of PBHs are the 21 cm absorption lines (see e.g., \cite{2012RPPh...75h6901P, Villanueva-Domingo:2021vbi, Liu:2022iyy} for reviews of the observable and \cite{Clark:2018ghm, Cang:2021owu, Villanueva-Domingo:2021cgh, Yang:2021idt} for PBH-related discussions). These lines are generated whenever a neutral hydrogen atom undergoes a spin-flip transition and are therefore a very important tracker of the neutral hydrogen distribution across space and time. The probability of this transition to happen is proportional to the relative abundance of the two spin levels, which in turn depends on what is known as the spin temperature $T_S$. Since $T_S$ is determined by the CMB and gas temperature, any process that affects the latter inevitably modifies also the 21 cm signal.

This logic has been applied to constrain several beyond-$\Lambda$CDM models (see e.g., \cite{Villanueva-Domingo:2021vbi} and references therein for a recent overview), and here we focus on the case of PBH evaporation \cite{Clark:2018ghm}. As extensively explained in the references, the relation between energy injection and modified 21 cm signal is dictated by the same equations discussed in the previous section in the context of CMB anisotropies. This similarity is qualitatively confirmed, for instance, in Fig. 4 of the reference, where the same $f_{\rm PBH}\propto M^3$ proportionality is shown for the 21 cm constraints. Therefore, in analogy to the previous section also in the 21 cm case we can simply recast the existing bounds of \cite{Clark:2018ghm} to account for the role of $\varepsilon$.

\begin{figure*}[t]
    \centering
    \includegraphics[width=\columnwidth]{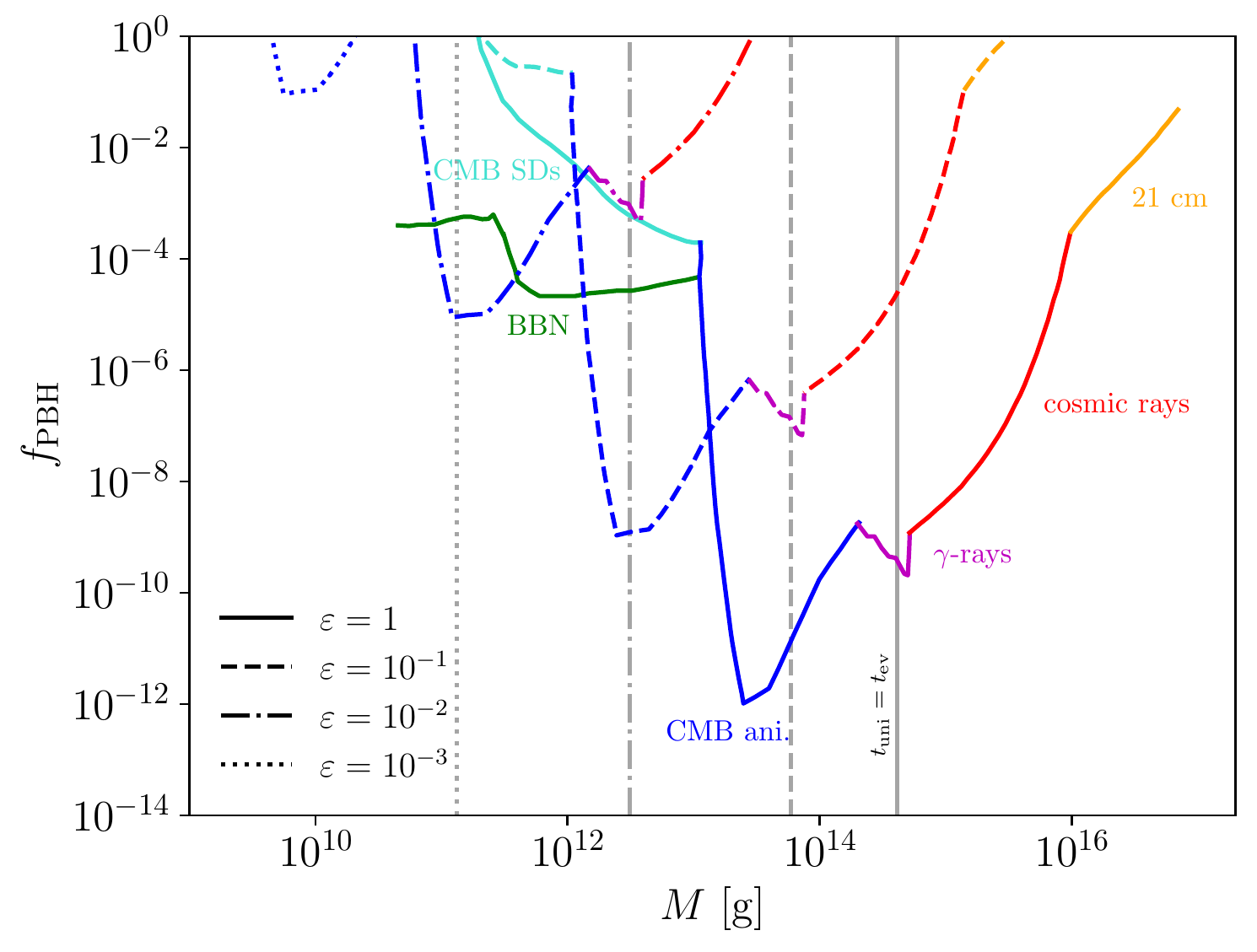}
    \includegraphics[width=\columnwidth]{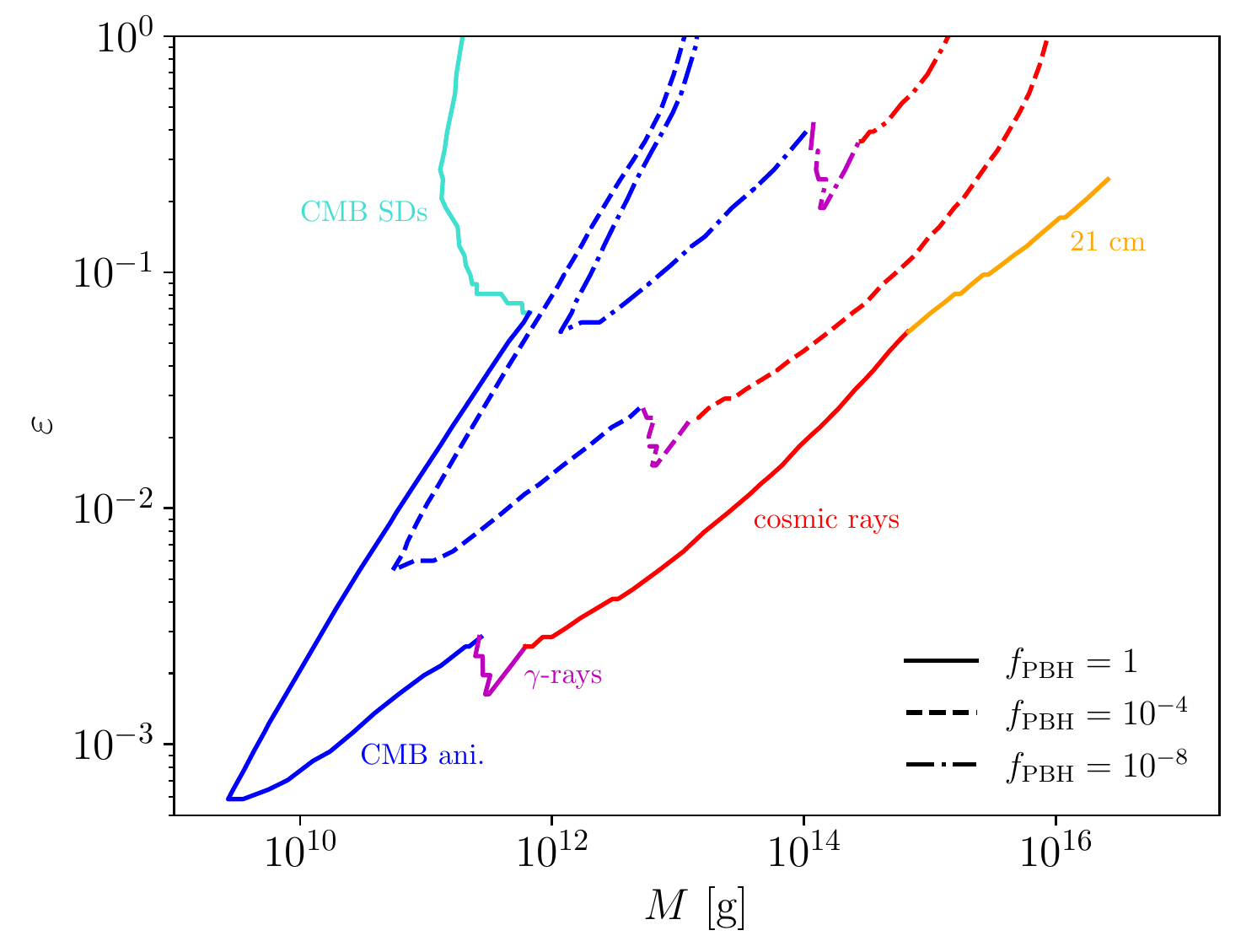}
    \caption{\textit{Left panel:} Cosmological and astrophysical constraints on the fractional PBH abundance as a function of the PBH mass for different values of the quasi-extremality parameter $\varepsilon$. The vertical gray lines represent the PBH masses whose lifetimes correspond to the age of the universe (with the same line styles). The BBN constraints are only shown for reference in the $\varepsilon=1$ case and are not rescaled for the other values of $\varepsilon$ for the reasons explained in the main text. \textit{Right panel:} Same as in the left panel but on the $\varepsilon-M$ plane for different values of $f_{\rm PBH}$.}
    \label{fig: bounds}
\end{figure*}

\subsection{Diffuse $\gamma$-ray background}\label{sec: gamma_ray}
Next we move to constraints of astrophysical origin, i.e., focusing on the evaporation of PBHs in the local environment. Firstly, we consider the case of the diffuse extra-galactic $\gamma$-ray background. The idea in this context is then to consider the observed $\gamma$-ray fluxes, observed by e.g., Fermi LAT \cite{Fermi-LAT:2014ryh} and HAWC \cite{Harding:2019tez}, and to impose the condition that the flux of photons emitted from the cosmological PBH population does not exceed this limit. This exercise has been performed in e.g., \cite{Carr2010New, Arbey:2019vqx, Chen:2021ngo, Keith:2022sow} (see Fig. 5 of \cite{Carr2010New}) including a number of observations and found that this probe is particularly constraining for PBH masses around $10^{15}$ g. The constraints from the galactic flux have been computed in \cite{2009A&A...502...37L, Carr:2016hva} and most recently in \cite{Laha:2020ivk}, but turn out to be subdominant with respect to the aforementioned 21 cm constraints in the relevant mass range and will therefore be neglected here.

Since these extra-galactic constraints depend on the flux of photons emitted from the PBHs, i.e.,
\begin{align}
    \phi_\gamma= \frac{1}{4\pi}    \int  L_{\gamma} \, n_{\rm PBH}\,  \text{d}t \propto \frac{f_{\rm PBH}}{t_{\rm ev}}\,,
\end{align}
where $L_\gamma$ is the emitted luminosity in form of photons and $n_{\rm PBH}$ is the PBH number density, also in this case the constraints have a power-law dependence on ${\varepsilon^4/(1+\varepsilon)^6}$, which allows for a straightforward rescaling of the aforementioned bounds derived in \cite{Carr2010New}.

\subsection{Cosmic rays}\label{sec: cosmic_rays}
A similar discussion can be also carried out for galactic cosmic rays, such as electrons and positrons \cite{macgibbon1991cosmic, carr1998cosmic, Barrau:1999sk}. The observational difficulty in this case is, however, that low-energy charged particles are significantly affected by the heliosphere of the sun and this limits the amount of information that can be extracted from the data. This problem has been overcome with the exit from the heliosphere of the Voyager 1 spacecraft \cite{doi:10.1126/science.1236408} and now it is therefore possible to combine Voyager 1 \cite{2016ApJ...831...18C} and \text{AMS-02}~\cite{2014PhRvL.113l1102A} data to constrain the cosmic ray flux over an energy range between a few MeV and hundreds of~GeV.

These data sets have been employed by \cite{Boudaud:2018hqb} to bound the PBH abundance in the mass range between $5\times 10^{14}-3\times10^{16}$ g, where they are also the most constraining to date. As for the $\gamma$-rays, also in this case the limits rely on the definition of the flux of particles from the PBHs, so that the same $\varepsilon$ rescaling applies here as well.

\subsection{Diffuse high-energy neutrino background}\label{sec: neutrino}
Finally,  the observation of the neutrino flux at facilities such as IceCube \cite{aartsen2017icecube} and Super-Kamiokande \cite{FUKUDA2003418} would enable us to constrain the PBH abundance in the local environment. However, so far the current sensitivity of these experiments has not been able to set competitive bounds with respect to the aforementioned ones \cite{020PhRvL.125j1101D2}. We will therefore not consider these observations in the following discussion, but point them out as a promising avenue for the future.

\subsection{Comment on non-evaporation constraints}
On top of the constraints that can be imposed on PBHs via their Hawking evaporation, also other bounds can be derived via e.g., eventual gravitational effects, astrophysical arguments as well as the accretion of matter onto the BHs (see e.g., \cite{Carr2020Primordial} for a recent review). Typically, all of these limits apply to masses well above $10^{20}$ g, where the lifetime of the PBHs is much larger than the age of the universe. In this regime, the effect of Hawking evaporation is greatly suppressed and the mass loss rate is negligible. As a consequence, these constraints are insensitive to the quasi-extremal origin of the PBHs and are not modified by the value of $\epsilon$.

A possible exception would be the limits due to the assumed formation mechanism. In fact, as mentioned above the production of qPBHs would require very specific characteristics that could affect the evolution of the universe aside from the consequent Hawking evaporation. Nevertheless, since these bounds would be strongly model dependent, we do not consider them any further in this work.

\section{Results}\label{sec: results}
The collection of the cosmological and astrophysical constraints on the PBH abundance discussed in the previous section is summarized in the left panel of Fig. \ref{fig: bounds}. There, we also display the BBN constraints derived in \cite{Carr2010New, Acharya2019CMB} for reference (solid green line), although they are not rescaled as the others and are only to be relied upon for the $\epsilon=1$ case. We remark, however, that the conclusions drawn below do not depend on this limitation of the analysis.

In the figure, the solid lines represent the cases with $\varepsilon=1$, while dashed, dashed-dotted and dotted lines refer respectively to the $\varepsilon=0.1, \, 0.01$ and 0.001 cases. As expected, the constraints are significantly suppressed and shifted to lower masses the lower the value of $\varepsilon$. In fact, as argued in the previous section the upper bound on the PBH abundance relaxes roughly proportionally to $\varepsilon^4/(1+\varepsilon)^6$ both vertically and horizontally, which in turns means that the largest mass allowed by evaporation constrains for $f_{\rm PBH}=1$ reduces to approximately $3\times10^{15}$ g for $\varepsilon=0.1$, to $3\times 10^{13}$ g for $\varepsilon=0.01$ and all PBH masses are allowed for $\varepsilon\lesssim 10^{-3}$.

We also show as vertical lines the PBH masses whose lifetime would correspond to the age of the universe (with the same line style as above). This sets the threshold above which the PBHs are still present in the universe today (or, alternatively, below which they are already evaporated). While in the $\varepsilon=1$ this corresponds to approximately $4.1\times10^{14}$ g, this value scales as in Eq.~\eqref{eq: t_ev}, i.e., approximately as $(\varepsilon^4/(1+\varepsilon)^6)^{1/3}$. This means that for $\varepsilon\sim 10^{-3}$ even PBHs as light as $\sim10^{11}$ g would survive until today.

The combination of these two conclusions, i.e., that qPBHs with $\varepsilon\leq10^{-3}$ can match the correct DM abundance and that they would still be present today, opens the door to the interesting possibility that such light qPBHs could be the DM. Of course, this will need to be developed in the context of more refined qPBH models, but can still act as a useful (conservative) benchmark for such scenarios.

Interestingly, for relatively small values of $\varepsilon$, the bounds presented in the left panel of Fig. \ref{fig: bounds} can be approximated to be on the parameter combination $f_{\rm PBH}\,\epsilon^4$ (neglecting the dependence on $f(T)$ and on the second order $\varepsilon$ term, see Eq.~\eqref{eq: E_prop_PBH}). This allows for a very simple, order-of-magnitude reinterpretation of the constraints for any value of $\varepsilon$. Furthermore, it also allows us to present the limits in the $\varepsilon-M$ plane for fixed values of $f_{\rm PBH}$, which can be useful for realistic models where the qPBHs are predicted to be a given sub-component of the DM. We perform this exercise (with the exact dependence of Eq.~\eqref{eq: E_prop_PBH}) in the right panel of Fig. \ref{fig: bounds} for the representative cases of $f_{\rm PBH}=1,\, 10^{-4}, \, 10^{-8}$. The figure confirms the aforementioned discussion.

\section{Summary and discussion}\label{sec: conclusions}
The analysis carried out here focuses on PBHs in the mass range between $\sim10^{10}-10^{17}$ g. The allowed abundance of such PBHs is mostly constrained by the impact of their evaporation on cosmological and astrophysical observables such as the CMB, the 21 cm lines and cosmic rays. Nevertheless, these stringent limits are derived assuming non-spinning, neutral (i.e., Schwarzschild) BHs, a scenario that maximizes the evaporation efficiency. If one assumes instead that the BHs are e.g., charged, spinning or even living in a higher-dimensional space, their evaporation temperature decreases, and consequently so does their luminosity. This approach can be pushed to the limit where the evaporation stops completely, leading to what are known as extremal BHs.

In this work we consider so-called quasi-extremal PBHs and show that indeed the assumption of quasi-extremality can greatly suppress the aforementioned constraints on the PBH evaporation. Concretely, we analyse the case of a general and conservative scenario where the degree of quasi-extremality is captured by a model-independent parameter $\varepsilon$, which we assume to be constant for simplicity. In the context of charged BHs, for instance, this quasi-extremality parameter would be defined as $\varepsilon^2=1-Q^2/M^2$, where $Q$ and $M$ represent charge and mass of the PBH, respectively. As a result, we find \textit{i)} that all constraints vanish for $\varepsilon\lesssim10^{-3}$ and \textit{ii)} that for these values of $\varepsilon$ all PBHs with masses larger than $\sim10^{11}$ g would still be present today. The combination of these conclusions implies that such light qPBHs are a viable DM candidate.

However, the question of observability remains. In fact, given the dependencies on $\varepsilon$ of the constraints discussed in Sec.~\ref{sec: observables} we do not expect that upcoming experiments such as CMB-S4 \cite{Abazajian2016CMB, Abazajian2019CMB} and SKA \cite{2013ExA....36..235M} (see e.g.,~\cite{Lucca2019Synergy, Mena:2019nhm} for related forecasts) would be able to significantly change the current picture since qPBHs with $\varepsilon\lesssim10^{-3}$ would still largely evade them. Furthermore, even if a survey did observe a signature compatible with the energy injection following the evaporation of PBHs, it would be impossible to disentangle the case of a Schwarzschild PBH population from that of a more abundant population of lighter qPBHs. Therefore, cosmological and astrophysical probes testing Hawking evaporation are a priori not sensitive enough to uniquely prove the existence of qPBHs and complementary observations would become fundamental.

For instance, while microscopic PBHs might make up for most of the DM if they are quasi-extremal, there may also be a high mass tail, which would provide a unique gravitational wave (GW) signature observable by future observatories such as LISA \cite{2017arXiv170200786A}. In fact, the degree of quasi-extremality has an impact on the GW signature in the case of a merger and values of $1-a$ (and, similarly, of $\varepsilon$) as small as $10^{-9}$ may be detectable in the waveform measurable by LISA for extreme mass ratio merger events~\cite{Burke:2020vvk}. Such BH-BH interactions might also leave unique signatures in the early universe, as would be the case for opposite charge BH encounters, although we leave a more accurate investigation of this possibility for future work. Another avenue to disentangle Schwarzschild and qPBHs in a potential cosmological observation is to determine the PBH mass independently, which can be achieved by e.g., gravitational direct detection~\cite{Carney:2019pza} and other direct detection techniques~\cite{Lehmann:2019zgt}.

In summary, in this work we have shown that light, quasi-extremal PBHs can be the DM and argued that with the help of complementary GW observations, an accurate determination of their characteristics might be within reach. The results found here are general and conservative, and should be taken as the basis for future, model-specific studies.

\section*{Acknowledgements}
We thank Marco Hufnagel, Vasilis Niarchos, Achilleas Porfyriadis and Christopher Rosen for very useful discussions. We also thank Andrew Kovachik and the McMaster University theory group for their helpful feedback on the manuscript. ML is supported by an F.R.S.-FNRS fellowship.

\appendix
\section{More details on the properties of four-dimensional and higher-dimensional black holes}\label{app: det_d}

In this appendix we collect some useful formulae that pertain to the properties of 4D as well as higher-dimensional Schwarzschild BHs. 

\subsection{4D Schwarzschild black holes}

The Schwarzschild radius is 
\be
r_s={2GM\over c^2}=2.95 {M\over M_{\odot}}~\text{km}\,,
\ee
where the Planck mass is given by
\be
M_P=\sqrt{\hbar c\over G}\simeq 1.2 {10^{19}\over c^2}~\text{GeV}\simeq 2.2 \times 10^{-8}~\text{kg}\,.
\ee
The Hawking temperature is given by
\be
T_H={\hbar c^3\over 8\pi k~GM}\simeq 2.5\times 10^{21}\left({1 ~{\rm gr}\over M}\right)~{\rm eV}\,.
\ee

The decay rate due to Hawking radiation to massless constituents is given by the Stefan-Boltzmann-like formula 
\be
{dM\over dt}=-4\pi \sigma r_s^2~T_H^4=-{\hbar c^6\over 30\cdot 8^3\pi G^2}{1\over M^2}\,,
\label{A1}\ee
where 
\be
\sigma={\pi^2\over 60}{k^4\over c^2\hbar^3}
\ee
is the standard Stefan-Boltzmann coefficient. This formula is also valid for massive particles emitted, provided their mass $m c^2\ll T_H$. For the Standard Model (SM) of particle physics plus Gravity, this means photons and gravitons,  For $M\gg 2.5\times 10^{18}$ g we can neglect therefore the emission of other SM particles. For $5\times 10^{12}$ g $\ll M \ll 2.5\times 10^{18}$ g, one should also include the three SM neutrinos. For $2.5 \times 10^{10}$ g $\ll M \ll 5\times 10^{12}$ g one should include electron emission, and so on.

Solving Eq. (\ref{A1}) we obtain for the evolution of the mass
\be
M(t)=\left(M_0^3-{\hbar c^6\over 10\cdot 8^3\pi G^2}t\right)^{1\over 3}
\ee
and therefore the evaporation time $t_{\rm ev}$ is given by
\be
ct_{\rm ev}={10\cdot 8^3~G^2 M^3\over\hbar c^5}={640\over \hbar G}r_s^3=640{M_P^2\over \hbar^2} r_s^3\,.
\ee
The evaporation formulae above are assuming massless photons. Including gravitons doubles the rate. Grey-body factors are ignored as they are not important for standard Schwarzschild BHs as absorption cross sections are geometrical in the IR and suppressed in the UV regime.

\subsection{General dimension $d\geq 4$ }
We now move to $d$ spacetime dimensions with $d\geq 4$ and we set $\hbar=c=1$. In this case, the definitions of Schwarzschild radius and Hawking temperature can be generalized as
\be
r_s^{d-3}=2G_dM\quad \text{and}\quad kT_H={d-3\over 4\pi r_s}\,,
\label{A8}\ee
while the decay rate due to "massless" Hawking radiation is given by
\be
-{dM\over dt}=\Omega_{d-2}~r_s^{d-2}\sigma_{d}(kT_{H})^d=~{S_d\over r_s^2}
={S_{d}\over (2G_d M)^{2\over d-3}}
\label{A9}\ee
with
\be
\Omega_{d-2}\equiv {2\pi^{d-1\over 2}\over \Gamma\left[{d-1\over 2}\right]}\,, \quad S_d\equiv \Omega_{d-2}\sigma_d\left({d-3\over 4\pi}\right)^d
\ee
and $\sigma_d$ is the analogue of the Stefan-Boltzmann coefficient in $d$ dimensions.

Solving Eq. (\ref{A9}) we obtain
\be
(2G_d)^{2\over d-1}~M(t)=\left[\left(2G_d M_0^{d-1\over 2}\right)^{2\over d-3}-{d-1\over d-3}S_d t\right]^{d-3\over d-1}
\ee
and
\be
t^{(d)}_{\rm ev}={d-3\over d-1}{\left(2G_d M_0^{d-1\over 2}\right)^{2\over d-3}\over S_d}\,.
\ee
Consider now the $d-4\equiv n$ extra dimensions to be compactified on $T^n$ with all radii equal to $R$ for simplicity.
In that case the 4D, $G$, and $d$-dimensional Newton constants, $G_d$, are related as
\be
G_d=G~(2\pi R)^{d-4}\,.
\ee
We may rewrite the $d$-dimensional evaporation time in this case as
\be
t^{(d)}_{evap}(M)={d-3\over d-1}{\left(2G (2\pi R)^{d-4} M^{d-1\over 2}\right)^{2\over d-3}\over S_d}
\ee
from which it follows that
\begin{align}
\nonumber {t^{(d)}_{\rm ev}(M)\over t^{(4)}_{\rm ev}(M)} & =3{(d-3)\over (d-1)}{S_4\over S_d}\left({\pi R\over GM}\right)^{2{(d-4)\over (d-3)}}
\\ & =3{(d-3)\over (d-1)}{S_4\over S_d}\left({2\pi R\over r_s}\right)^{2{(d-4)}}\,.
\label{life}\end{align}
Simple physical arguments indicate that when the BH is much smaller in size than $R$, i.e., with
\be
r_s\ll R\,,
\ee
then it behaves as a $d$-dimensional BH. From Eq. (\ref{A8}) we obtain that $kT_H\gg {1\over R}$ and therefore the BH can radiate all Kaluza-Klein (KK) modes of the graviton and other fields. Its lifetime from Eq. (\ref{life}) is much longer than a 4D BH of the same mass. Moreover, if $r_s\ll R$ then during the evaporation process, the horizon radius becomes smaller and smaller and the whole evaporation process happens in the $d$-dimensional regime.

On the other hand for the BH to be semi-classical, we must have
\be
G_d r_s^{2-d}\ll 1\ar {G\over (2\pi R)^2}\left({2 \pi R\over r_s}\right)^{d-2}\ll 1\,,
\ee
which implies the inequalities
\be
1\ll {t^{(d)}_{\rm ev}(M)\over t^{(4)}_{\rm ev}(M)}\ll \left({(2\pi R)^2\over G}\right)^{2{d-4\over d-2}}\to \left({2 \pi c M_P R\over \hbar}\right)^{4{d-4\over d-2}}\,.
\ee
In the extreme case, $R\simeq 1$ $\mu$m, we obtain  
$${c\over \hbar} M_P R~\simeq~  10^{28}\;.$$

If on the other hand we have a BH with a horizon radius $r_s\gg R$, in this case the BH behaves as 4D BH. The temperature is much smaller than the KK mass scale and none of the KK states can be emitted. While it is evaporating, it will start doing so by using the 4D formula, but as its horizon radius becomes smaller than $R$, then it starts evaporating as a $d$-dimensional BH. The transition mass $M_*$ is given by $r_s=R$,
\be
2 G~M_*=R\ar M_*={M_PR\over 2}M_P
\ee
with 
\be
M_P R\lesssim 10^{28}\,.
\ee
In the extreme case of $R= 1~\mu$m, we obtain
\be
M_*\simeq  10^{23}~{\rm gr}\,.
\ee 

Simplifying, we assume that the evaporation process happens as 4D, until $M$ reduces to $M_*$ and as higher $d$ when $M<M_*$, and we obtain,
\be
\small
M(t)= \left\{ \begin{array}{lll}
    \displaystyle
    \left[M^3-{3S_4 t\over (2G)^2}\right]^{1\over 3}
    ,&\quad &0<t\leq t_*,\\ \\
    \displaystyle
    \left[M_*^{d-1\over d-3}-{d-1\over d-3}{S_d (t-t_*)\over (2G_d)^{2\over d-3}}\right]^{d-3\over d-1}
    ,&\quad&t_*<t\leq T(M)
\end{array}\right.
\label{1}
\ee
where $t_*$ is the time for the BH to reach the transition mass $M_*$
\be
M_Pt_*={2\over S_4}\left({M^3\over M_P^3}-{M_*^3\over M_P^3}\right)
\ee
and $T(M)$ is the total evaporation time
\be
T(M)=t_*+{d-3\over d-1}{(2G_d)^{2\over d-3}M_*^{d-1\over d-3}\over S_d}\,.
\ee
When $M\gg M_*$, then
\begin{align}
\nonumber {t_*\over T-t_*} & \simeq {(d-1)\over 3(d-3)}{S_d\over S_4}\left({2GM\over R}\right)^3\\ & ={(d-1)\over 3(d-3)}{S_d\over S_4}\left({r_s\over R}\right)^{3(d-3)}\gg 1
\end{align}
and essentially, the decay time is given by $t_*$ which is approximately equal to the 4D evaporation time
\be
T(M)\simeq t_*\simeq t_{evap}\equiv {(2G)^2 M^3\over 3S_4}\,.
\ee
If on the other hand the mass $M<M_{*}$ then the evaporation is higher-dimensional and in that case
\be
T(M)=t^{(d)}_{\rm ev}\equiv  {d-3\over d-1}{(2G_d)^{2\over d-3}M^{d-1\over d-3}\over S_d}
\label{A27}\ee
We conclude that ``large" BHs $M\gg M_*$ have a 4D decay time, while ``small" BHs, $M\ll M_*$ have a higher dimensional decay time given in Eq. (\ref{A27}). For a small BH, taking the ratio of Eq. (\ref{A27}) with the 4D one we obtain
\be
{t^{(d)}_{\rm ev}\over t_*}\simeq {t^{(d)}_{\rm ev}\over t^{(4)}_{\rm ev}}=3{d-3\over d-1}{S_4\over S_d}\left({2\pi R\over r_s}\right)^{2(d-4)}\,.
\ee
Therefore small BHs have $r_s\ll R$ and are relatively long-lived compared to 4D Schwarzschild BHs.

We can define an effective extremality parameter $\epsilon_{\rm eff}$ for small higher-dimensional BHs as
\begin{equation}
    \varepsilon_{\rm eff}^{-4}={3{(d-3)\over (d-1)}{S_4\over S_d}\left({2\pi R\over r_s}\right)^{2{(d-4)}}}
    \label{A29}\end{equation}
by comparing it with 4D RN BHs (see Eq.~\eqref{eq: t_ev}).

\bibliography{bibliography}{}

\end{document}